\documentclass[longbibliography,twocolumn,pra,aps,superscriptaddress,showpacs,amsmath,amssymb,floatfix]{revtex4-1}
\usepackage{graphicx}
\usepackage{amssymb}
\usepackage{amsmath}
\usepackage{epsfig}
\usepackage{color}
\usepackage{tabu}
\usepackage{mathtools}
\usepackage[colorlinks,linkcolor=blue,anchorcolor=blue,citecolor=blue,urlcolor=blue]{hyperref}
\usepackage{physics}
\usepackage{float}
\usepackage{diagbox}

\usepackage{url}

\usepackage{listings}
\definecolor{dkgreen}{rgb}{0,0.5,0.7}
\definecolor{gray}{rgb}{0.5,0.5,0.5}
\definecolor{mauve}{rgb}{0.58,0,0.82}
\begin{document}

\title{Quafu-Qcover: Explore Combinatorial Optimization Problems on Cloud-based Quantum Computers}

\author{BAQIS Quafu Group}

\date{\today}

\begin{abstract}
We present Quafu-Qcover, an open-source cloud-based software package designed for combinatorial optimization problems that support both quantum simulators and hardware backends.
Quafu-Qcover provides a standardized and complete workflow for solving combinatorial optimization problems using the Quantum Approximate Optimization Algorithm (QAOA). It enables the automatic modeling of the original problem as a quadratic unconstrained binary optimization (QUBO) model and corresponding Ising model, which can be further transformed into a weight graph. The core of Qcover relies on a graph decomposition-based classical algorithm, which enables obtaining the optimal parameters for the shallow QAOA circuit more efficiently.
Quafu-Qcover includes a specialized compiler that translates QAOA circuits into physical quantum circuits capable of execution on Quafu cloud quantum computers. Compared to a general-purpose compiler, ours generates shorter circuit depths while also possessing better speed performance. The Qcover compiler can establish a library of qubits coupling substructures in real time based on the updated calibration data of the superconducting quantum devices, ensuring that the task is executed on physical qubits with higher fidelity. 
The Quafu-Qcover allows us to retrieve quantum computer sampling result information at any time using task ID, enabling asynchronous processing.
Besides, it includes modules for result preprocessing and visualization, allowing for an intuitive display of the solution to combinatorial optimization problems. We hope that Quafu-Qcover can serve as a guiding example for how to explore application problems on the Quafu cloud quantum computers.
\end{abstract}

\maketitle

\section{ introduction}
Quantum computing is a new computing paradigm, which is intended to take advantage of superposition and entanglement of quantum systems to accelerate computation. The basic information processing unit of quantum computing is quantum bit, i.e., qubit. Different from classical bits that only be one of the states $0$ or $1$, qubits can be prepared in the superposition of both states $|0\rangle+|1\rangle$, which implies the exponential representation capability of state space. For example, the state space of only $300$ qubits is already comparable with the total number of atoms in our universe. The core of quantum algorithms is to extract the desired solution from the exponential state space by designing suitable quantum operations on qubits. There exists some quantum algorithms already be proved theoretically that has the ability to surpass their classical counterpart on certain issues, e.g., Shor, Grover, and HHL algorithms \cite{RevModPhys.82.1,montanaro2016quantum}. The Shor algorithm can factor a large integer in polynomial time, while the classical algorithm takes exponential time \cite{Shor1994,Shor1999}. The Grover algorithm can search for a specific item in an unsorted database with only $O(\sqrt{N})$ time complexity, which has square acceleration compared to the existing classical algorithm \cite{PhysRevLett.79.325}. The HHL algorithm can solve some specific linear equation problems in polynomial time, while classical computers need exponential time to solve these problems \cite{HHL}.   

Due to the rapid development of quantum technologies in the past decade, kinds of physical platforms have been constructed to demonstrate quantum computing tasks \cite{gulde_implementation_2003,benhelm_towards_2008,taylor_fault-tolerant_2005,PhysRevLett.104.040502,clarke_superconducting_2008,barends_digitized_2016,arute_quantum_2019,PhysRevLett.127.180501,doi:10.1126/science.1142892,PhysRevLett.117.210502,PhysRevLett.123.250503,doi:10.1126/science.abe8770,kane_silicon-based_1998,PhysRevLett.89.017901,he_two-qubit_2019}. Among which the superconducting quantum platform has attracted a lot of attention because of its scalability and high fidelity of gates operation. In 2019, Google first demonstrated the so-called quantum supremacy on the problem of random circuit sampling by using its $53-$qubit superconducting quantum processor ``Sycamore" \cite{arute_quantum_2019}. The record was updated in 2021 by USTC Pan group with a $66-$qubit superconducting quantum processor ``Zuchongzhi II"  on the same problem \cite{PhysRevLett.127.180501}. Meanwhile, superconducting-based quantum computing cloud platforms have evolved rapidly in the past few years, e.g., IBM Quantum. To date, IBM has launched $20+$ quantum processors on the cloud ranging from $5$ qubits to a maximum of $127$ qubits \cite{PhysRevA.94.012314,wang_16qubit_2018,PhysRevLett.122.080504,HuangIBM53,ball2021first}. The IBM Quantum has already served millions of registered users and billions of quantum circuit tasks have been executed by the cloud.

However, currently all quantum computers lack fault-tolerance. Achieving fault-tolerant quantum computing would require millions of physical qubits with high gate operation fidelity, which is far beyond current technical capabilities. For the foreseeable future, we will be and remain in the era of ``Noisy Intermediate Scale Quantum"(NISQ) for a long time. During the NISQ stage, quantum processors contain tens to hundreds of physical qubits but suffer from noise that no efficient error corrections can be implemented \cite{Preskill2018quantumcomputingin}. Due to the constraint of not good enough qubits quality and gates operations, the depth of executable quantum circuits will be very limited, i.e., only shallow quantum circuits can be efficiently executed. The most urgent and challenging thing in the NISQ era is to demonstrate quantum benefits in application scenarios.  

There are some fields, such as molecular simulation, material design, and combinatorial optimization, which are believed to be suitable for exploration in the NISQ era \cite{RevModPhys.94.015004}. Among these, combinatorial optimization attracts a lot of attention and study, as it is involved in a wide range of industrial production scenarios, including logistics, finance, and telecommunications, to name a few. Combinatorial optimization problems, which are NP-hard, imply that no classical algorithms of polynomial complexity exist as the problem scale increases. Quantum algorithms are expected to perform better in finding near-optimal solutions.
In 2014, E. Farhi {\it et al} proposed the Quantum Approximation Optimization Algorithm (QAOA) for solving combinatorial optimization problems \cite{farhi2014quantum}.
QAOA is a variational algorithm, which makes it suitable for implementation on NISQ processors. Its core idea is to convert the optimization problems into finding ground state solutions of the Ising Hamiltonian, which can be searched using the variational method. To date, many variants of QAOA have been proposed, such as Recursive QAOA \cite{PhysRevLett.125.260505}, Warm-Start QAOA \cite{Egger2021warmstartingquantum}, and Reinforcement Learning-assisted QAOA \cite{patel2022reinforcement}, among others.
Although QAOA has some limitations \cite{farhi2020quantum,PRXQuantum.2.030312,PhysRevLett.124.090504,stilck_franca_limitations_2021}, for certain instances, QAOA-related algorithms have the potential to outperform classical algorithms \cite{PhysRevLett.125.260505,9605292,PhysRevA.104.052419}.

The Cloud-based quantum computing service makes quantum resources available to a wider audience by enabling users to remotely access quantum computers and execute quantum programs.  As more and more people use and explore quantum computing in the cloud, there is a significant need to present guiding examples of full-stack application software to users. On the one hand, it helps beginners explore quantum computing in specific application fields without a professional background in quantum computing. On the other hand, it provides potential high-level quantum developers with a reference example to develop their own cloud-based quantum software.
Therefore, in this work, we present Quafu-Qcover, an open-source quantum software that helps users explore combinatorial optimization problems with the QAOA algorithm. The Quafu-Qcover supports direct calls to quantum computers via the Quafu quantum computing cloud platform \cite{quafu}. The basic workflow of Quafu-Qcover consists of four main stages. 
Firstly, combinatorial optimization problems can be effectively modeled as Quadratic Unconstrained Binary Optimization (QUBO) model, which can be equivalently transformed into the Ising Hamiltonian form through suitable variable transformations. 
Secondly, the QAOA circuit is built according to the Ising Hamiltonian with undetermined parameters. The optimal parameters are determined by the graph decomposition method with each sub-graph corresponding one-to-one with smaller QAOA circuits that can be executed on quantum hardware or simulators \cite{zhuang2021efficient}. 
Thirdly, the QAOA circuit is compiled into an executable quantum circuit on the Quafu quantum computers through an efficient compiler. It should be noted that the determination of optimal parameters and circuit compilation can be realized in parallel. 
Finally, the compiled circuit is sent to Quafu quantum computers to execute sampling, and users can check the task status at any time using the task ID. Upon completing the sampling task, the Quafu-Qcover returns the results of problems in a visualization format and the bit strings of the QUBO model.

It should be emphasized that to execute quantum algorithms on practical quantum computers, the quantum programs have to be converted into hardware-executable quantum gate sequences or pulse sequences through a quantum compiler. The quantum compilation is the most important bridge connecting quantum programs and quantum hardware. 
In the present NISQ era, the primary objective of quantum compilation is to optimize quantum circuits by reducing the circuit depth and the number of two-qubit gates, while considering the topology and characteristics of the underlying hardware structure. Although it has been shown that finding the optimal quantum circuit is usually an NP-hard problem \cite{botea2018complexity,10.1145/3168822,10.1145/3569052.3578928}, in practice, we only need to search for sub-optimal quantum circuits within a reasonable time.
For some circuits with special structures, such as the QAOA circuit, specific compiling strategies can outperform general-purpose compilers \cite{10.1145/3470496.3527394,9218558,
9251960,Weidenfeller2022scalingofquantum,jin2021structured}. Inspired by previous works and in consideration of the topology and characteristics of the Quafu quantum processor, we design a specific compiler for Quafu-Qcover that exhibits superior performance compared to general-purpose compilers like Qiskit.

The remainder of this paper is organized as follows: In Sec. II, we provide a comprehensive overview of the fundamental background of the QUBO model and the QAOA. In Sec. III, we present the Quafu-Qcover software framework and elaborate on the functionalities of each module in detail. Subsequently, in Sec. IV, we illustrate the practical application of Quafu-Qcover by presenting a specific combinatorial optimization problem and outlining the step-by-step process. Finally, in Sec. V, we provide a comprehensive summary of the entire paper and engage in a discussion regarding potential avenues for future research.

\section{ QUBO and QAOA Basics} 
Combinatorial optimization problems play a significant role in various applications \cite{glover_quantum_2022}, including maximum independent set (MIS) problems, maximum cut (MaxCut) problems, graph coloring problems, task allocation problems, etc. These diverse combinatorial optimization problems can be effectively modeled using a universal framework known as the QUBO form.
The QUBO model has gained significant attention in the field of quantum annealing and serves as a key component in experiments conducted with quantum annealers developed by D-Wave Systems \cite{glover_quantum_2022,neven2009nips}. Additionally, it serves as the foundational framework for the QAOA, which operates on gate-based quantum computers.
The objective function of the QUBO model can be expressed as follows:
\begin{equation}
f(x) = \sum_{i,j} Q_{ij}x_i x_j =  x^T Q x, \quad x_i \in \{0,1\},
\label{QUBO}
\end{equation}
where $x \in \{0,1\}^n$ is a vector of binary variables, and $Q \in \mathbb{R}^{n\times n}$ is a real square matrix.
For combinatorial optimization problems, we need to optimize the objective function and find the binary vector $x^*$ that minimizes $f(x)$.
That is
\begin{equation}
x^{*}=\min _{x \in \{0,1\}^n} f(x).
\label{x*}
\end{equation}
By defining $s_i = 2x_i - 1$, the QUBO model is inherently connected to the classical Ising model,
\begin{equation}
H=\sum_{i, j} J_{i j} s_i s_j+\sum_i h_i s_i \quad s_i \in \{-1,+1\}.
\end{equation}
The problem of finding the optimal solution $x^*$ as described in Eq. \ref{x*} is equivalent to the task of determining the ground state of the Ising model. In other words, it involves identifying the configuration of binary labels $\{-1, +1\}$ that minimizes the energy of the classical spin Hamiltonian.

Based on the aforementioned observations, it is also natural to establish a mapping between combinatorial optimization problems and the quantum Ising model. The quantum Ising Hamiltonian is represented as follows:
\begin{equation}
H_C= \sum_{i,j} J_{i j} Z_i Z_j + \sum_i h_i Z_i,
\end{equation}
where $Z_i$ are Pauli-Z operators with eigenvalues $z_i=\pm 1$, $J_{i j}$ represent the interactions between two spins and $h_i$ denote the biases associated with each individual spin.
In the computational basis, the vectors $|z\rangle$ are represented by $z\equiv z_1z_2\cdots z_n$, and the Hamiltonian $H_C$ is diagonal. Consequently, the objective function is defined as $C(z)=\langle z| H_C |z\rangle$. The optimal solution to the combinatorial optimization problem requires identifying the bitstrings $z$ that minimize or maximize the value of $C(z)$. 
However, the majority of optimization problems are known to be NP-hard, which means that finding the optimal solution becomes increasingly difficult as the problem size grows. Therefore, it becomes almost impractical to obtain the exact optimal solution for such problems. 

To address this challenge, the QAOA was proposed by Farhi et al \cite{farhi2014quantum}. It is a variational quantum algorithm that is applicable to current NISQ processors. The QAOA usually introduces the following mixer Hamiltonian,
\begin{equation}
H_B=\sum_{i=1}^N X_i,
\end{equation}
where $X_i$ are the Pauli-X operators. The QAOA aims to provide sub-optimal solutions to the objective Hamiltonian $H_C$ by iteratively applying the objective Hamiltonian $H_C$ and the mixer Hamiltonian $H_B$ with optimized parameters. 
The initial state of the QAOA circuit is commonly prepared as a uniform superposition state of all $n$ qubits. This state can be represented as follows:
\begin{equation}
|S\rangle = |+\rangle^{\otimes n}=\frac{1}{\sqrt{2^n}} \sum_z|z\rangle,
\end{equation}
which is the eigenstate of $H_{B}$.
Subsequently, the unitary operators $U_C(\gamma)=e^{-i \gamma H_C}$ and $U_B(\beta)=e^{-i \beta H_B}$ are sequentially applied to the initial state in an alternating manner, resulting in the generation of a variational state, denoted as
\begin{equation}
|\overrightarrow{\gamma}, \overrightarrow{\beta} \rangle = U_B(\beta_p) U_C(\gamma_p) \cdots U_B(\beta_1) U_C(\gamma_1)|+\rangle^{\otimes n},
\label{eq-vwf}
\end{equation}
This variational state is parameterized by $2p$ real variational parameters $\gamma_i$ and $\beta_i$ $(i = 1,2,\ldots,p)$, which control the angles of rotation associated with the objective Hamiltonian $H_C$ and the mixer Hamiltonian $H_B$, respectively. 
The expectation value of the objective Hamiltonian $H_C$ in this variational state is given by:
\begin{equation}
E_p(\vec{\gamma}, \vec{\beta})=\langle\vec{\gamma}, \vec{\beta}|H_C| \vec{\gamma}, \vec{\beta}\rangle.
\label{eq-Ep}
\end{equation}
To initialize the parameters $\vec{\gamma}$ and $\vec{\beta}$, we can use either some random guesses or employ warm starting methods \cite{Egger2021warmstartingquantum,PhysRevA.104.052419}. The latter provides strategies for initializing the parameters in a way that improves the convergence and performance of the QAOA algorithm.
Next, classical optimizers are utilized to search for the optimal parameters $\{\vec{\gamma}_{\text{opt}}, \vec{\beta}_{\text{opt}}\}$ that minimize or maximize $E_p$. Based on the optimized parameters and Eq. \ref{eq-vwf}, the corresponding quantum circuit can be prepared to obtain the best approximation solution state $|\overrightarrow{\gamma}_{\text{opt}}, \overrightarrow{\beta}_{\text{opt}} \rangle$ using a quantum computer.
By measuring in the computational basis, a bitstring $z$ is obtained, and the value of $C(z)$ is evaluated. Repeating the measurements with the same optimized quantum circuit, the bitstring $z^*$ that maximizes or minimizes $C(z)$ corresponds to an approximate optimal solution for the combinatorial optimization problem.

\begin{figure*} 
\centering
\includegraphics[width=0.98\textwidth]{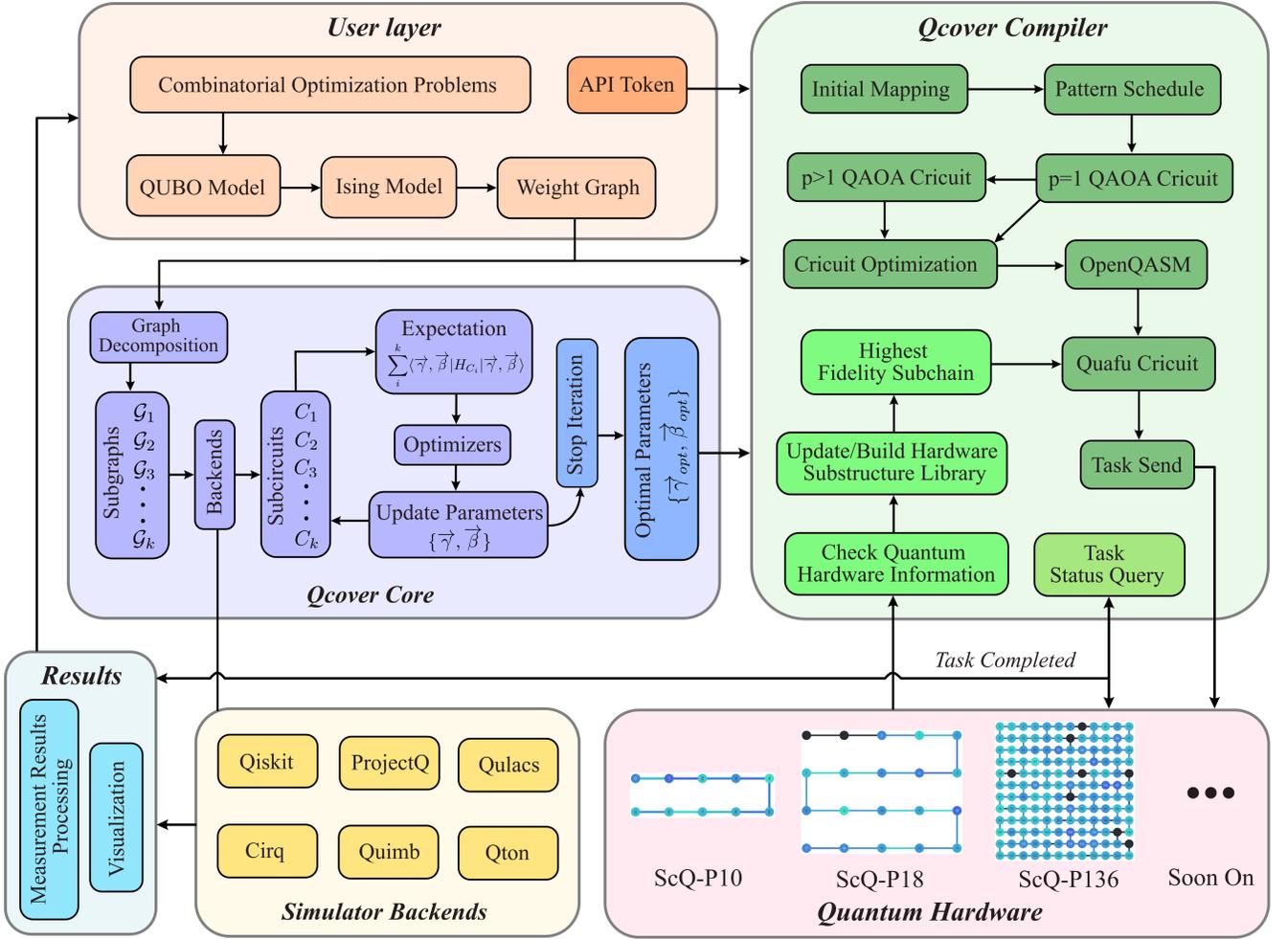}
\caption{Quafu-Qcover software framework. At the {\it user level} module, users are required to input the combinatorial optimization problem they wish to solve and provide parameters such as the API token of Quafu. The {\it Qcover core} module is an essential component that utilizes the graph decomposition algorithm to determine optimal parameters for the QAOA circuit. The {\it Qcover compiler} module is used to compile the QAOA circuit into a physical quantum circuit that can be executed on Quafu hardware. The {\it simulator backends} consists of various open-source quantum programming packages. The {\it quantum hardware} encompasses the superconducting quantum chips currently accessible on the Quafu quantum computing cloud platform. The {\it results} module is responsible for processing the sampling results obtained from Quafu and presenting them to the users in a visual format.}
\label{fig-fig1}
\end{figure*}

\section{Software Architecture}
We now introduce the software architecture of Quafu-Qcover. The Quafu-Qcover is primarily composed of the following modules: user layer, core module, compiler module, quantum simulator backends, quantum hardware backends, and results preprocessing module. 

\subsection{User layer}
The user layer is an interaction module between users and software. For a given combination optimization problem, we first need to model it as a QUBO form, then convert the QUBO to the corresponding Ising Hamiltonian form.  The Ising form itself cannot be directly dealt with and has to be mapped onto a weight graph $G(V, E)$ as input for the next layer.
The spins $Z_i$ in the Ising Hamiltonian correspond to the vertices $V_{i}$ in graph $G$, with $h_{i}$ being the weight of the vertices. The $Z_iZ_j$ terms correspond to the edges $E_{ij}$ in $G$, with $J_{ij}$ being the weights of the edges. As for users, they only need to know how to model optimization problems as a QUBO form. All relevant quantum components have been packaged in the software, and users can use them directly without a background in quantum theory. In addition, we have pre-written tens of typical optimization problems in the {\it Qcover.applications} module of Quafu-Qcover that can be transformed into corresponding weight graphs for computation. To call quantum computers in the Quafu cloud, users need to provide their own API token, which can be easily obtained on the Quafu website \cite{quafu}.

\subsection{Core module}
The core module of Quafu-Qcover focuses on finding the optimal parameters for the QAOA circuit using a graph decomposition algorithm.
The fundamental idea is to utilize graph decomposition and graph-to-circuit mapping methods to partition a large-scale QAOA instance into numerous independent smaller instances, which can be processed in parallel using quantum simulators or hardware \cite{zhuang2021efficient}.
For most of the optimization problems, our algorithm exhibits a linear scaling of the running time with respect to the number of qubits. Most of the widely-used quantum simulators are compatible with Quafu-Qcover. We have integrated several popular simulator software packages as our backends, including Qiskit \cite{Qiskit}, ProjectQ \cite{projectq}, Qulacs \cite{qulacsfast}, $t|ket\rangle$ \cite{Sivarajah_2021}, and Quimb \cite{gray2018quimb}. Additionally, we have also developed our own simulator called Qton \cite{qton}, which is also included in the software suite.

Quafu-Qcover includes several classic optimizers for updating parameters in QAOA circuits. These optimizers encompass gradient descent-based algorithms and other optimization algorithms such as COBYLA, L-BFGS-B, SHGO, SLSQP, and SPSA. In addition, some heuristic algorithms suitable for multi-layer QAOA circuits are also included, for example, Fourier and Interp algorithms \cite{PhysRevX.10.021067}.

The core module operates according to the following basic workflow. Initially, employing the graph decomposition algorithm, the input weight graph $\mathcal{G}$ is decomposed into a sequence of small weighted subgraphs ${\mathcal{G}_1, \mathcal{G}_2, ..., \mathcal{G}_k}$.
Subsequently, each small weighted subgraph $\mathcal{G}_i$ is mapped to a corresponding QAOA subcircuit $C_i$ based on the selected software backend framework.
The subcircuits $C_i$ correspond to different Ising Hamiltonians $H_{C_i}$, each of which includes distinct terms from $H_C$ without overlapping. To evaluate these subcircuits, we compute their expectations independently and in parallel using the selected simulator or hardware. Then, the expectation value $E_p$ of $H_{C}$ is given by 
\begin{equation}
E_p(\vec{\gamma}, \vec{\beta})= \sum_i^k \langle\vec{\gamma}, \vec{\beta}|H_{C_i}| \vec{\gamma}, \vec{\beta}\rangle.
\label{eq-EpHCi}
\end{equation}
Finally, a selected classical optimizer is employed to optimize the parameters. The iteration is halted and the optimal parameters are acquired once the expectation value attains the desired accuracy or reaches a predetermined threshold.

\subsection{Compiler module}
Another pivotal module within Quafu-Qcover is a specialized compiler specifically designed for QAOA circuits. The Qcover compiler possesses the capability to transform the weight graph of any given input problem into an executable QAOA circuit suitable for quantum processors. Subsequently, this circuit is sent to the quantum hardware backend for computation. Next, we introduce the Qcover compiler.

\begin{figure*} 
\centering
\includegraphics[width=0.98\textwidth]{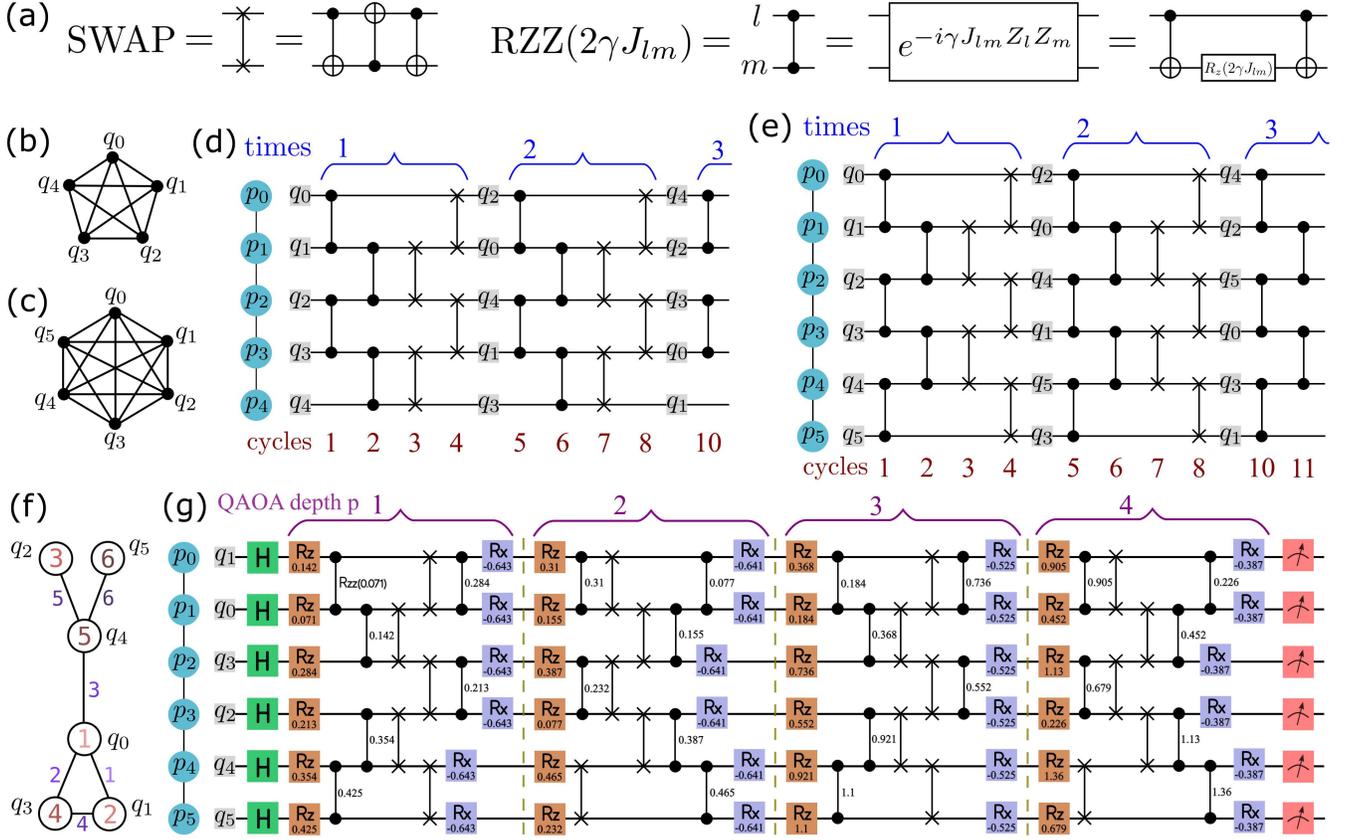}
\caption{ (a) Decomposition rules for SWAP gates and RZZ gates. (b) and (c) are fully connected weight graphs with 5 nodes and 6 nodes, respectively. (d) and (e) correspond to the compilation templates of fully connected weight graphs with 5 nodes and 6 nodes on a one-dimensional chip, respectively. (f) A specific example of a weight graph for the Ising model. The values labeled on the nodes and edges indicate the weights. (g) The weight graph illustrated in (f) is compiled into a quantum circuit that satisfies the one-dimensional qubits coupling structure using the Qcover compiler. }
\label{fig-fig2}
\end{figure*}

\begin{figure*} 
\centering
\includegraphics[width=0.98\textwidth]{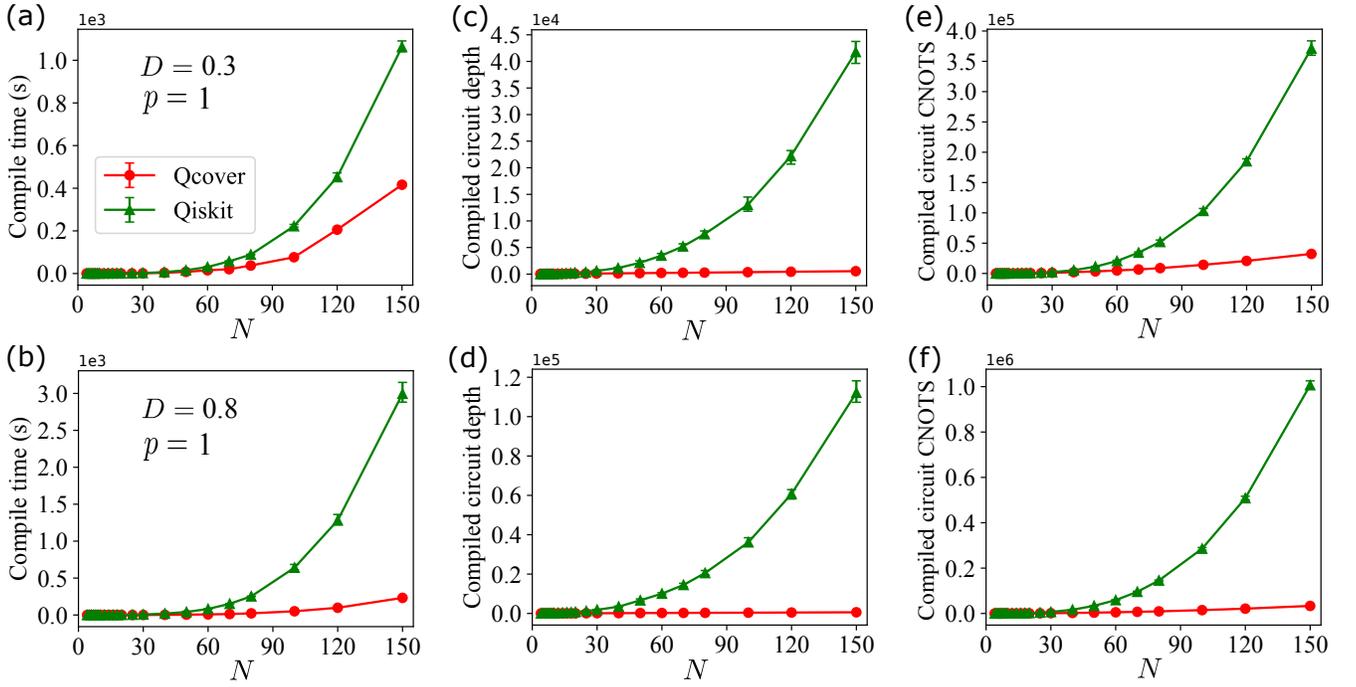}
\caption{Performance comparison of the Qcover compiler and the Qiskit compiler. (a)-(b), (c)-(d), and (e)-(f) represent the variation of compilation time, compiled circuit depth, and the number of compiled circuit CNOT gates with the number of weight graph nodes. The edge density of the weight graphs in the upper and lower panels are 0.3 and 0.8, respectively. The Qiskit compiler selects the highest compilation optimization level. The green triangle dots and red dots indicate the mean results obtained from compiling 20 weight maps randomly generated using the Qiskit compiler and the Qcover compiler, respectively.}
\label{fig-fig3}
\end{figure*}

\begin{figure*} 
\centering
\includegraphics[width=0.98\textwidth]{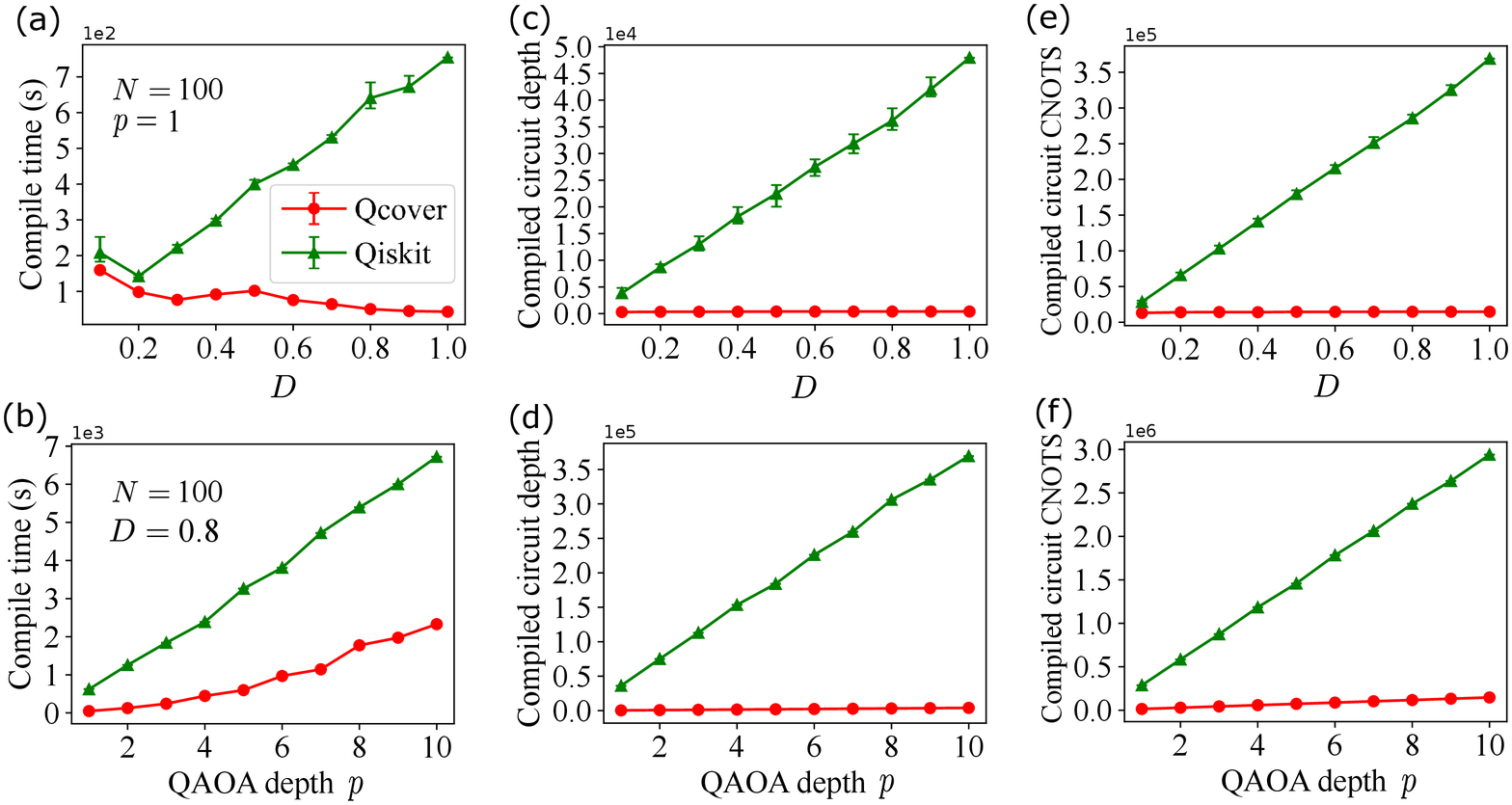}
\caption{In the upper panel, the variation of compiler performance with the density of weight graph edges is depicted for a fixed weight graph node number $N=100$ and QAOA circuit depth $p=1$. Notably, the performance of the Qcover compiler outperforms Qiskit by one to two orders of magnitude, and the performance improvement is more substantial with increasing edge density. In the lower panel, for a fixed weight graph node number $N=100$ and weight graph edge density $D=0.8$, the performance of the compiler is shown as a function of QAOA circuit depth $p$. The circuit depth and the number of CNOT gates compiled by Qcover are dozens of times smaller than those compiled by Qiskit, and this performance improvement remains almost unaffected by the QAOA depth.}
\label{fig-fig4}
\end{figure*}

In order to obtain the variational wave function, it is necessary to construct the unitary evolution operators $U_B(\beta_i)$ and $U_C(\gamma_i)$ into quantum circuits. Since the terms in $H_B$ consist of commuting single-qubit operators, $U_B(\beta)$ can be expressed as the product of each term, as shown below:
\begin{equation}
U_B(\gamma) = e^{-i\beta H_B} = \prod_{l} e^{-i\beta X_l},
\label{eq-UB}
\end{equation}
where each operator $e^{-i\beta X_l}$ corresponds to a single qubit RX$(2\beta)$ rotation gate.
For $U_C(\gamma)$, since all terms in $H_C$ are diagonal in the computational basis, these terms commute with each other. Consequently, $U_C(\gamma)$ can be expressed as
\begin{equation}
U_C(\gamma) = e^{-i\gamma H_C} = \prod_{lm} e^{-i\gamma J_{lm} Z_lZ_m} \prod_{l} e^{-i\gamma h_{l} Z_l},
\label{eq-UC}
\end{equation}
where the individual operator $e^{-i\gamma J_{lm}Z_lZ_m}$ and $e^{-i\gamma h_{l}Z_l}$ correspond to the two-qubits RZZ($2\gamma J_{lm}$) gate and the single-qubit RZ($2\gamma h_{l}$) rotation gate, respectively.

As previously stated, in the case of a Hamiltonian consisting exclusively of Pauli-Z operators, the terms commute with each other,  enabling the unitary operator $U_C$ to be expressed precisely as the product form in Eq. \ref{eq-UC}. 
However, when the Hamiltonian contains arbitrary Pauli operators, we can employ the Trotter theorem to approximate the unitary operator $U_C$ \cite{lloyd1996universal,doi:10.1063}.
In this paper, we focus on the objective Hamiltonian that exclusively consists of Pauli-Z operators, which is applicable to most combinatorial optimization problems.

Next, the compilation of the original QAOA circuit into a physical quantum circuit capable of execution on a real quantum processor is required. Various compilation methods have been developed for QAOA circuits. In this study, we primarily focus on a structured method introduced in Ref. \cite{jin2021structured}. This method demonstrates that when executing the full connection weight graph on a one-dimensional quantum processor with nearest-neighbor coupling, the resulting QAOA circuit exhibits a structured form.
Assuming a mapping of $N$ logical qubits ($q_i$) onto a one-dimensional quantum processor with $N$ physical qubits ($p_i$), the QAOA circuit is constructed by interleaving RZZ gates and SWAP gates in the following manner ($i=0,1,2,\ldots$) \cite{jin2021structured}:

(1) Execute RZZ gate between all $p_{2i}$ and $p_{2i+1}$ physical qubits;

(2) Execute RZZ gate between all $p_{2i+1}$ and $p_{2i+2}$ physical qubits;

(3) Execute SWAP gate between all $p_{2i+1}$ and $p_{2i+2}$ physical qubits;

(4) Execute SWAP gate between all $p_{2i}$ and $p_{2i+1}$ physical qubits.

For a complete graph with $N$ nodes, the aforementioned four steps need to be repeated $[N/2]$ times. Specifically, if $N$ is even, $2N-2$ cycles are executed; if $N$ is odd, $2N-1$ cycles are executed.
For any incomplete graph, the template circuit can be used by inserting the required RZZ gates and SWAP gates accordingly. It should be noted that different initial mappings may impact the circuit depth of the incomplete graph, but they do not affect the circuit depth of the complete graph. Nonetheless, this compilation approach remains effective for any weighted graph. Since this compilation pattern is fixed, the position of the inserted RZZ gate can be predicted. Thus, once the initial qubit mapping is determined, the cycle of each RZZ gate in the circuit can be specified. This allows us to identify the optimal mapping that minimizes the overall quantum circuit depth.

Although it is possible to determine the circuit depth for a given initial mapping, finding the optimal initial mapping requires exploring $N!$ possibilities. As the size of the graph grows, it becomes infeasible to explore every potential initial mapping.
To maximize the likelihood of obtaining the optimal initial mapping, we employ a heuristic search algorithm inspired by the approach described in Ref. \cite{jin2021structured}. This algorithm relies on two key observations:
(1) When we map a new logical qubit $q_i$ to a physical qubit $p_j$, we can use this pattern to determine the last cycle that $q_i$ is executed among all other currently mapped qubits. (2) The vertices in the weight graph that exhibit higher connectivity exert a more significant influence on the resulting circuit depth. Next, we introduce the heuristic search algorithm for finding the optimal initial mapping. In this algorithm, each node within the search tree represents a mapping of a logical qubit to a physical qubit.
The algorithm flow is as follows:

{\noindent Step 1. Sort the vertices of the input problem graph in descending order based on their degrees and store them in the list $Sq$. Vertices with higher degrees are mapped first. 
\\
Step 2. Level 0 of the search tree comprises a single root node, initialized as empty, indicating the absence of any mappings. Subsequently, we generate child nodes for the root node by mapping the highest degree logical qubit $q=Sq[1]$ to each unmapped physical qubit. As a result, level 1 encompasses $N$ nodes, each with a cost of zero. Furthermore, we add the mapped logical qubit $q=Sq[1]$ into the initially empty list $Mq$ for future reference. 
\\
Step 3. The child nodes of level 2 are generated from the parent nodes of level 1. We map the second-highest degree logical qubit $q=Sq[2]$ to each unmapped physical qubit $p$, resulting in level 2 containing $N(N-1)$ nodes. The cost of each node is determined as follows:
\begin{equation}
cost_{q \rightarrow p} = \max_{m\in M_q \land  m\leftrightarrow q} (\text{ExeR}(q, m)),
\end{equation}
where $m\leftrightarrow q$ indicates that the mapped logical qubit $m$ and the current logical qubit $q$ are connected in the input weight graph $\mathcal{G}$, and $\text{ExeR}(q, m)$ represents the cycle at which the RZZ$(q,m)$ gate is scheduled in the pattern.
In order to avoid the excessive breadth of the search tree, we impose a truncation on each level. It should be explained that although node A costs less than node B under the current mapping, it does not guarantee that the node generated by node A will cost less than the node generated by node B when adding subsequent mapping. Therefore, the truncation here does not depend on the cost of the nodes, but for nodes with the same cost, we keep $B_{\max}$ nodes and expand them.  Since the cost corresponds to the cycle of the RZZ gate in the circuit, there can be a maximum of $2N$ distinct costs. As a result, the maximum breadth of the search tree is $2NB_{\max}$. Here, the maximum cycle is not expanded, and the constant term is disregarded. This approach allows us to manage the breadth of the search tree effectively.
Unless otherwise specified, we set $B_{\max}=5$. A larger value of $B_{\max}$ may yield a more optimal initial mapping, but it also increases the search time. Upon completion of this step, we record the cost associated with the expanded node and include the mapped logical qubit $q=Sq[2]$ in the list $Mq$.
\\
Step 4. Proceeding to step 3, we generate subsequent levels of the search tree. For instance, the child nodes of level 3 are generated from the nodes of level 2. This implies that we map the third-highest degree logical qubit $q=Sq[3]$ to each unmapped physical qubit $p$. It should be noted that without truncation of the search tree breadth, level 3 will contain $N(N-1)(N-2)$ nodes. The algorithm concludes when all logical qubits are successfully mapped. 
Ultimately, we obtain a search tree with a maximum breadth of $2NB_{\max}$ and a depth of $N$. Each path within the search tree corresponds to a distinct initial mapping. By considering the associated costs, we identify the mapping that minimizes the overall circuit depth, thereby serving as the optimal initial mapping.
}

The structured compilation method described above is applicable to the QAOA circuit with $p=1$. According to Eq. \ref{eq-UB} and Eq. \ref{eq-UC}, we generalize this method to any integer $p$.
For the compilation of the QAOA circuit with $p>1$, all the odd layers remain the same as the $p=1$ circuit, while all the even layers are the reverse of the $p=1$ circuit. An example is illustrated in \ref{fig-fig2}(g). It should be noted that the optimization parameters of each layer are different and require modification.

The quantum circuit obtained above includes RZZ gates and SWAP gates, and cannot be directly executed on superconducting hardware. Therefore, we need to decompose them into CNOT gates and RZ gates according to the rules shown in Fig. \ref{fig-fig2}(a). Subsequently, we further optimize the circuit, primarily involving the cancellation of adjacent CNOT gates with identical control and target qubits, and aligning all gates to the left to execute them at the earliest possible opportunity. This optimization leads to a reduction in the number of gates and circuit depth. The optimized circuit is then converted into OpenQASM format. With these steps, we have successfully completed the compilation of any QAOA circuit on a one-dimensional superconducting chip. This method is also applicable to chips with arbitrary structures. We only need to identify the one-dimensional substructure chain of the chip and remap the qubits onto this subchain. Additionally, our compiler can be transplanted to other qubit systems, such as trapped ion systems, by decomposing the RZZ gates and SWAP gates into the basic gate set supported by the specific quantum hardware.

In addition, considering that we are currently in the NISQ era with limited qubit fidelity, it becomes crucial to take into account the fidelity information of qubits in quantum chips during the execution of quantum programs. 
The Qcover compiler incorporates the quantum hardware subchain library building module, which utilizes a heuristic algorithm to identify subchains with high overall fidelity in chips featuring arbitrary coupling structures. The term "overall fidelity" in this context denotes the product of the fidelities of all two qubits within the subchain.
Due to the influence of the external environment, the fidelity of qubits is not static, thus requiring frequent calibration of the qubits. The Qcover compiler module can update the subchain library in real time according to the quantum hardware information. 
The subchain library is represented as a dictionary-type data structure. Each entry in the dictionary consists of a key-value pair, where the key represents the number of qubits contained in the subchain, and the value represents the physical qubit coupling structure for each subchain. Note that a single key in the dictionary may correspond to multiple subchains. These subchains are sorted in descending order based on their overall fidelity.
When creating a circuit using the \textit{quafu} class, the subchain with the highest overall fidelity is selected from the library based on the required number of qubits. Subsequently, the logical qubits are mapped to the physical qubits of the selected subchain in a sequential manner.
Finally, we utilize the \textit{from\_openqasm}() function within the \textit{pyquafu} framework to convert OpenQASM code into \textit{quafu} class circuits. Subsequently, the task will be sent to the Quafu quantum cloud platform for computation using the \textit{send}() function.

\subsection{Compiler performance benchmark}
Performance indicators for evaluating the effectiveness of a compiler encompass several factors, including the compilation time, circuit depth, and gate count of the resulting physical quantum circuit. These metrics serve as crucial measures to assess the efficiency and quality of the compilation process.
To assess and compare the performance of different compilers, we conducted experiments using both the Qiskit compiler and our Qcover compiler on a personal computer. We evaluated the performance of these two compilers across varying parameters, including the weight graph edge density $D = \text{edges}/C_N^2$, the number of nodes $N$, and the QAOA circuit depth $p$.

In Fig. \ref{fig-fig3}, we present the outcomes of compiling QAOA circuits with a depth of $p=1$ using both the Qiskit compiler and our Qcover compiler. The upper panel corresponds to a random weight graph with an edge density of $D=0.3$, while the lower panel corresponds to a weight graph with a higher edge density of $D=0.8$. The left, middle, and right panels illustrate the compilation time, compiled circuit depth, and CNOT gate count as a function of the number of nodes, respectively. For a given number of nodes $N$ and edge density $D$, we randomly generated 20 weight graphs. Subsequently, we compiled these weight graphs using the two compilers individually, and the compiled results were averaged to provide a more comprehensive evaluation. 
In Fig. \ref{fig-fig3} (a) and (b), it is evident that the Qcover compiler exhibits significantly superior performance compared to the Qiskit compiler. The compilation time of Qiskit is several to dozens of times longer than that of Qcover.
In Fig. \ref{fig-fig3} (c) and (d), it can be observed that as the number of nodes increases, the depth of the Qiskit compiled circuit grows exponentially, whereas the depth of our Qcover compiled circuit increases linearly.
In Fig. \ref{fig-fig3} (e) and (f), we present the disparity in the count of CNOT gates between the compiled circuits of the two compilers. It is evident that the Qcover compiled circuit requires substantially fewer CNOT gates compared to the Qiskit compiled circuit. 
As the number of nodes increases, the count of CNOT gates in the Qiskit compiled circuit exhibits exponential growth, whereas the count of CNOT gates in our Qcover compiled circuit shows polynomial growth.

In Fig. \ref{fig-fig4}, we present the performance of the two compilers under other various parameters.
The upper panel displays the variation of each performance metric with the edge density of the weight graph when $N=100$ and $p=1$.
We observe that for any edge density, the performance of Qcover surpasses that of Qiskit, and the improvement in Qcover performance becomes more pronounced as the edge density increases. As the edge density of the weight graph increases, the compilation time of Qiskit exhibits nearly linear growth, except for very small values of $D$. In contrast, the compilation time of Qcover demonstrates a decreasing trend. This can be attributed to the adoption of the template compilation pattern in Qcover, wherein the time required to find the optimal mapping decreases as the edge density approaches $D=1$. 
The compiled circuit depth and the count of CNOT gates for Qiskit demonstrate a linear growth as the edge density of the weight graph increases. However, for Qcover, these quantities show an initial rapid increase followed by a tendency to saturate as the edge density increases.
The lower panel in Fig. \ref{fig-fig4} illustrates the performance variation of the Qcover and Qiskit compilers with respect to the QAOA depth $p$, when the weight graph has parameters $N=100$ and $D=0.8$. 
It should be noted that these results are obtained from a single random weight graph. For different values of $p$, the compilation performance of Qcover is superior to that of Qiskit, and the degree of performance improvement is nearly independent of $p$. The primary reason for this observation is that as the QAOA depth $p$ increases, both Qiskit and Qcover demonstrate a linear growth in terms of compiled circuit depth and the count of CNOT gates. Additionally, it is worth mentioning that for each layer of the QAOA circuit, we have the capability to perform parallel optimal parameter configuration and circuit optimization. This implies that the compilation time will primarily depend on the QAOA circuit with $p=1$. However, in this study, the compilation process was performed sequentially. In the later iterations of Quafu-Qcover, parallelization and parametric QAOA circuit compilation will be supported. In conclusion, our comprehensive evaluation of various parameters and performance metrics confirms the significantly enhanced performance of the Qcover compiler compared to Qiskit. This further underscores the importance of designing dedicated compilers for specific classes of application algorithms.

\subsection{Backends}
Quafu-Qcover integrates various classical simulator backends and quantum hardware backends. The classical simulator backends include Qiskit \cite{Qiskit}, ProjectQ \cite{projectq}, Qulacs \cite{qulacsfast}, $t|ket\rangle$ \cite{Sivarajah_2021}, Quimb \cite{gray2018quimb}, and Qton \cite{qton}, as mentioned before. The specific details and instructions for using these simulators can be found in their respective software documentation, and will not be reiterated here. In this section, our focus is on introducing the quantum hardware backends provided by the Quafu quantum computing cloud platform.

The Quafu cloud platform currently offers three superconducting quantum chips as its quantum computing backend. Specifically, the platform offers two one-dimensional architecture chips: the 10-qubit chip (ScQ-P10) and the 18-qubit chip (ScQ-P18). Additionally, there is a two-dimensional architecture chip called the  ScQ-P136.
For the ScQ-P10 chip, the average energy relaxation time ($T_1$) and the average phase coherence time ($T_2$) of its qubits are 30.878$\mu$s and 3.877$\mu$s, respectively. The average fidelity of the two-qubit gates on this chip stands at 95.9\%.
In the case of the ScQ-P18 chip, the $T_1$ and $T_2$ values are determined to be 35.492$\mu$s and 3.285$\mu$s, respectively, while the average fidelity of its two-qubit gates reaches 95.7\%.
Regarding the ScQ-P136 chip, the $T_1$ and $T_2$ values are measured at 8.585$\mu$s and 10.471$\mu$s, respectively. Moreover, the average fidelity of its two-qubit gates is 92.4\%. It should be noted that due to various environmental factors, regular recalibration is required for superconducting quantum chips, resulting in slight variations in the parameter information of the aforementioned chips.
In future versions of Quafu-Qcover, we aim to enhance the quantum backend by incorporating a more diverse range of systems, including ion trap and Rydberg atom systems, thereby offering a more comprehensive platform for quantum computing.

\section{Example demonstration}
In this section, we will outline a comprehensive workflow demonstrating the utilization of Quafu-Qcover for solving combinatorial optimization problems. 

In the first step, it is necessary to represent the combinatorial optimization problem as a corresponding weight graph. Within the {\it Qcover.applications} module, we offer two methods for constructing the weight graph.
One of the methods available is to construct a weight map from the QUBO matrix. This approach is more general, but it requires the user to manually convert the combinatorial optimization problem into a QUBO model. The implementation of this method can be found in the {\it common} class.
Furthermore, we propose another method to construct the weight graph from the original problem. Specifically, corresponding classes have been developed for various types of combinatorial optimization problems, enabling us to convert the original problem into a QUBO model, thus obtaining the weight graph. The Quafu-Qcover presently encompasses common optimization problems, including Maxcut, graph coloring, number partitioning, set packing, and quadratic assignment. For further details, please refer to the Quafu-Qcover project documentation. In future versions of Qcover, we plan to include more types of combinatorial optimization problems. The following are some example code snippets to obtain the weight graph from either the QUBO matrix or the original problem.

\begin{lstlisting}
from Qcover.applications import common, max_cut, graph_color, number_partition

# Graph data are described by the networkx python package
# Obtain the weight graph from the QUBO matrix
ising_mat = common.get_ising_matrix(qubo_mat)
weight_graph = common.get_weights_graph(ising_mat)

# Obtain the weight graph from the Maxcut problem
nodes = [0, 1, 2, 3, 4, 5]
edges = [(0,1),(1,2),(2,3),(3,4),(4,5),(0,4),(1,3)]
problem_graph = nx.Graph()
problem_graph.add_nodes_from(nodes)
problem_graph.add_edges_from(edges)
mc = max_cut.MaxCut(graph=problem_graph)
weight_graph, _ = mc.run()

# Obtain the weight graph from the graph coloring problem
gc = graph_color.GraphColoring(graph=problem_graph, color_num=3)
weight_graph, _ = gc.run()

# Obtain the weight graph from the number partition problem
nlist = [13, 7, 5, 2, 34, 21, 9, 45]  
np = number_partition.NumberPartition(number_list=nlist)
weight_graph, _ = np.run()
\end{lstlisting}

The second step involves utilizing the {\it Qcover.core} module to determine the optimal parameters for the QAOA circuit. This module provides a range of classical simulation backends and optimizers. In future versions of Qcover, an iterative approach will be introduced to directly identify the optimal parameters using quantum hardware. The following example code demonstrates the calculation of optimal parameters for the QAOA circuit.

\begin{lstlisting}
from Qcover.core import Qcover
from Qcover.backends import CircuitByQulacs
from Qcover.optimizers import COBYLA

# Set QAOA depth
p = 1   

# Select a simulation backend
bc = CircuitByQulacs()    

# Select an optimizer
optc = COBYLA(options={'tol': 1e-3, 'disp': True})  

# Instantiate the Qcover object
qc = Qcover(weight_graph, p=p, optimizer=optc, backend=bc)

# Run Qcover to find the optimal parameters
res = qc.run()
optimal_parameters = res['Optimal parameter value']
\end{lstlisting}

The third step entails compiling the QAOA circuit into a physical quantum circuit that can be executed by the Quafu quantum cloud platform based on the optimal parameters and weight graph. Subsequently, the compiled quantum circuit is sent to the cloud platform, which assigns a unique task ID number. The task status can be monitored using this ID number. Once the task is completed, the circuit sampling results can be obtained. Within the Qcover compilation package, we offer preprocessing and visualization methods for the sampling results. Currently, binary grouping problems, for example, Maxcut and number partitioning problems, can be visualized, while graph coloring problems visualization is also supported. The following code example demonstrates this step.

\begin{lstlisting}
from Qcover.compiler import CompilerForQAOA

# Set the API Token, select the quantum backend of the quafu cloud platform, and set the number of sampling
token = "Your API Token"
cloud_bc = 'ScQ-P18'
shots = 100   

# Instantiate the Qcover compiler object
qcover_compiler = CompilerForQAOA(weight_graph, p=p, optimal_params=optimal_parameters, apitoken=token, cloud_backend=cloud_bc)

# Compile the weight graph into the Quafu quantum circuit supported by the selected cloud backend and send the quantum circuit to the Quafu cloud platform to return the task ID number.
task_id = qcover_compiler.send(wait=False, shots=shots, task_name='MaxCut')

# If you choose wait=Ture, you have to wait for the result to return. If you choose wait=False, you can execute the following command to query the result status at any time, and the result will be returned when the task is completed. 
quafu_solver = qcover_compiler.task_status_query(task_id)
if quafu_solver:
    counts_energy = qcover_compiler.results_processing(quafu_solver)
    qcover_compiler.visualization(counts_energy, problem='MaxCut', solutions=2, problem_graph=problem_graph)
\end{lstlisting}

\begin{figure}[htp] 
\centering
\includegraphics[width=0.47\textwidth]{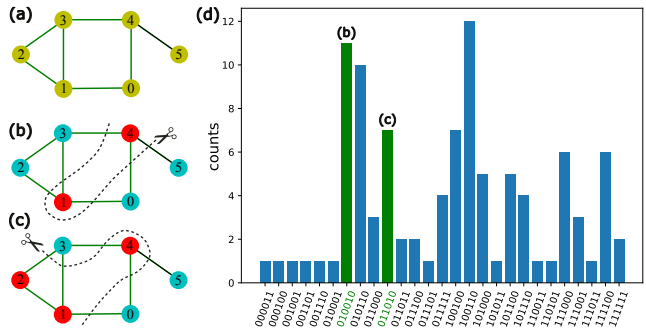}
\caption{(a) An example of the Maxcut problem. The nodes and edges are unweighted, or alternatively, it can be considered that the weights of the nodes and edges are all 1. (b) and (c) are two solutions given by the Quafu quantum computers. (d) Sampling results. The green highlights correspond to the two optimal solutions identified for (b) and (c).}
\label{fig-fig5}
\end{figure}

In Fig. \ref{fig-fig5}(a), we exhibit a specific case of the Maxcut problem, which was then submitted to the Quafu quantum cloud computing platform for computation, following the aforementioned steps. The sampling results obtained from the quantum processor are depicted in Fig. \ref{fig-fig5} (d), with the optimal solutions highlighted in green. These optimal solutions are further visualized in Fig. \ref{fig-fig5}(b) and (c).

\section{Conclusion}
Due to the limitation of quantum resources, the Qcover currently does not support the direct searching of optimal parameters with quantum hardware. The quantum chips are only called for sampling tasks in the last stage. At present, we are developing the runtime module, which supports sending a variational quantum circuit task. This module will facilitate the iterative search for optimal parameters of the QAOA circuit on a quantum computer.

In conclusion, we have introduced Quafu-Qcover, which is an open-source cloud-based quantum software. It enables users to remotely access the Quafu quantum computing cloud platform and use quantum computers for exploring combinatorial optimization problems. Quafu-Qcover utilizes a graph decomposition algorithm to efficiently determine the optimal parameters for QAOA circuits. In addition, it provides a customized high-performance QAOA circuit compiler that selects the qubits with the highest fidelity to perform tasks based on the fidelity information of the quantum chip at the backend of Quafu. The workflow of Quafu-Qcover is designed to be simple and user-friendly. The users only need to input the corresponding QUBO model or weight graph of the combinatorial optimization problem, utilize classical optimizers to find the optimal parameters for the QAOA circuit, and then send them to the Quafu quantum computing cloud platform asynchronously after compilation by the compiler.

{\bf Code available:} The corresponding software can be found at:https://github.com/BAQIS-Quantum/Qcover.

\section*{Acknowledgement}
This work is supported by the Beijing Academy of Quantum Information Sciences.



\bibliography{ref}

\newpage
\onecolumngrid

\vspace{1em}
\begin{flushleft}
{\small BAQIS Quafu Group}

\bigskip
{\small
\renewcommand{\author}[2]{#1$^\textrm{\scriptsize #2}$}
\renewcommand{\affiliation}[2]{$^\textrm{\scriptsize #1}$ #2 \\}

\newcommand{\corrauthora}[2]{#1$^{\textrm{\scriptsize #2}, \ddagger}$}
\newcommand{\corrauthorb}[2]{#1$^{\textrm{\scriptsize #2}, \mathsection}$}

\newcommand{\xBAQIS}{\affiliation{1}{Beijing Academy of Quantum Information Sciences, Beijing 100193, China}}

\newcommand{\xTsinghua}{\affiliation{2}{State Key Laboratory of Low Dimensional Quantum Physics, Department of Physics, Tsinghua University, Beijing, 100084, China}}

\newcommand{\xCAS}{\affiliation{3}{Institute of Physics, Chinese Academy of Sciences, Beijing 100190, China}}

\newcommand{\xFrontier}{\affiliation{4}{Frontier Science Center for Quantum Information, Beijing 100184, China}}

\newcommand{\xCASE}{\affiliation{5}{School of Physical Sciences, University of Chinese Academy of Sciences, Beijing 100190, China}}

\newcommand{\xCASF}{\affiliation{6}{CAS Center for Excellence in Topological Quantum Computation, UCAS, Beijing 100190, China}}



\newcommand{\BAQIS}{1}
\newcommand{\Tsinghua}{2}
\newcommand{\CAS}{3}
\newcommand{\Frontier}{4}
\newcommand{\CASE}{5}
\newcommand{\CASF}{6}

\author{Hong-Ze Xu}{\BAQIS},
\author{Wei-Feng Zhuang}{\BAQIS},
\author{Zheng-An Wang}{\BAQIS},
\author{Kai-Xuan Huang}{\BAQIS},
\author{Yun-Hao Shi}{\CAS, \!\CASE, \!\CASF},
\author{Tian-Ming Li}{\CAS, \!\CASE, \!\CASF},
\author{Chi-Tong Chen}{\CAS, \!\CASE, \!\CASF},
\author{Kai Xu}{\CAS, \!\BAQIS}
\author{Yu-Long Feng}{\BAQIS},
\author{Pei-Liu}{\BAQIS},
\author{Mo Chen}{\BAQIS}
\author{Shang-Shu Li}{\CAS, \!\CASE, \!\CASF},
\author{Zhi-Peng Yang}{\BAQIS},
\author{Chen Qian}{\BAQIS},
\author{Yu-Xin Jin}{\BAQIS}
\author{Yun-Heng Ma}{\BAQIS},
\author{Xiao Xiao}{\BAQIS},
\author{Peng Qian}{\BAQIS},
\author{Yanwu Gu}{\BAQIS},
\author{Xu-Dan Chai}{\BAQIS}
\author{Ya-Nan Pu}{\BAQIS},
\author{Yi-Peng Zhang}{\BAQIS},
\author{Shi-Jie Wei}{\BAQIS},
\author{Jin-Feng Zeng}{\BAQIS},
\author{Hang Li}{\BAQIS},
\author{Gui-Lu Long}{\Tsinghua, \!\BAQIS},
\author{Yirong Jin}{\BAQIS},
\author{Haifeng Yu}{\BAQIS},
\author{Heng Fan}{\CAS, \!\BAQIS, \!\CASE, \!\CASF},
\author{Dong E. Liu}{\Tsinghua, \!\BAQIS, \!\Frontier},
\corrauthorb{Meng-Jun Hu}{\BAQIS},

\bigskip

\xBAQIS
\xTsinghua
\xCAS
\xFrontier
\xCASE
\xCASF

{${}^\mathsection$ Corresponding author: humj@baqis.ac.cn}

}
\end{flushleft}

\end{document}